\documentclass[journal]{IEEEtran}

\def\^{\hat}
\def\~{\tilde}

\def\3h{{3\over 2}}

\def\v#1{{\bf#1}}

\def\H{{\v H}}

\def\J{{\v J}}

\def\eqn#1$${\eqno{{\rm #1}}$$}

\def\~{\tilde}

\def\^{\hat}

\def\vg#1{\mbox{{\boldmath${#1}$}}} 

\def\XXint#1#2#3{{\setbox0=\hbox{$#1{#2#3}{\int}$}
     \vcenter{\hbox{$#2#3$}}\kern-.5\wd0}}

\usepackage{graphicx}
\usepackage{epstopdf}
\input psfig.tex
\usepackage{multirow}
\usepackage{cite}

\usepackage{caption}
\usepackage{subcaption}
\usepackage{amsmath}
\usepackage{color}
\usepackage{fancyhdr}

\usepackage{float}
\allowdisplaybreaks[4]
\begin{document}

\title{ The Spectral Integral Method (SIM) for the Scattering from an Arbitrary Number of Circular PEC Cylinders}

\author{Qing Huo Liu$^*$, Siwei Wan, and Chunhui Zhu
\thanks{$^*$Corresponding author: Qing Huo Liu (e-mail: qhliu@duke.edu).
Qing Huo Liu and Siwei Wan are with the Department of Electrical and Computer Engineering, Duke University, Durham NC 27708-0291, USA. }
\thanks{Chunhui Zhu is with the Institute of Electromagnetics and Acoustics, Xiamen University, China}
}



\markboth{IEEE Transactions on Microwave Theory and Techniques} {Liu: SIM for Multiple Circular PEC Cylinders}

\maketitle
\thispagestyle{plain}
\begin{abstract}
We present an accurate spectral integral method (SIM) for the analyses of scattering from multiple circular perfect electric conductor (PEC) cylinders. It solves the coupled surface integral equations by using the Fourier series and addition theorem to decouple the system. The SIM has exponential convergence so that the error decreases exponentially with the sample density on the surfaces, and requires only about 2-3 points per wavelength (PPW) to reach engineering accuracy with less than 1\% error.  Numerical results demonstrate that the SIM is much more accurate and efficient than the method of moments (MoM), and thus can be potentially used as the exact radiation boundary condition in the finite element and spectral element methods.
\end{abstract}


\section{Introduction}
\PARstart{T}{he} scattering of multiple circular perfect electric conductor (PEC) cylinders is of significant scientific and engineering interest. In this work, we develop a fast spectral integral method (SIM) for simulating the scattering from an arbitrary number of non-overlapping circular PEC cylinders as shown in Figure 1. The radii and center positions of all cylinders can be arbitrary.
\graphicspath{ {.} }
\captionsetup[figure]{name={Fig.},labelsep=period}

\begin{figure}[htbp]
\centering
\begin{subfigure}[b]{0.45\textwidth}
         \centering
         \includegraphics[width=0.7\textwidth]{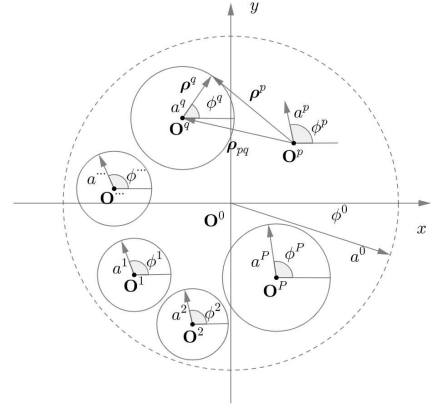}
         \label{fig1a}
     \end{subfigure}
\centering
\label{Fig.1}
\caption{Configuration of $P$ non-overlapping circular PEC cylinders of radius $a^p$ centered at ${\bf O}^p=(\rho_p,\phi_p)$ $(p=1, 2, \cdots, P)$.
Any point on the $p$-th cylinder can be conveniently written in its local polar coordinates as ${\vg\rho}=(a^p,\phi^p)$. The virtual outer circle with radius $a^0$ centered at the origin encloses all PEC cylinders. The $z$ direction is perpendicular to the page.}
\centering
\end{figure}

 Such a scattering problem is classical and can be treated by analytical \cite{Chew1995, 2Cylinders, NCylinders, Elsherbeni1987, Elsherbeni1992} and numerical methods \cite{GMT, Waterman1, Waterman2}.  The analytical solutions use cylindrical harmonics to expand the fields everywhere in terms of cylindrical harmonics and solve for the expansion coefficients.  Traditional numerical methods with the low-order method of moments (MoM) \cite{MoM}, finite element method and finite difference method can also be applied to solve this problem, but are time consuming, although the fast multipole method \cite{FMM} and multilevel fast multipole algorithm \cite{Song1995} have been applied successfully to accelerate the MoM.

Here we present a new perspective of this classical problem through fast surface integral equations for the unknown surface current density (rather than the fields everywhere). We develop a spectral integral method (SIM) for this scattering problem through the fast Fourier transform (FFT) algorithm and singularity subtraction to speed up the convergence of numerical solutions for the surface integral equations.  Although the SIM was first named in
\cite{JLiu2004}, its elemental ideas go back to the original work by Bojarski using FFT for a single cylinder \cite{Bojarski1984}, with the important singularity subtraction of the Green's function \cite{Hu1995}.  The SIM was further developed for objects in a layered medium \cite{Simsek2006}.  Both \cite{JLiu2004} and \cite{Simsek2006} use an iterative method to solve the system equation through FFT, thus the computational complexity is $O(KN\log N)$ for $N$ surface unknowns, where $K$ is the number of iterations.  The FFT has also been used for the scattering of multiple cylindrical layers \cite{Schuster1990}, although without treating the singularity of the Green's function and hence with slow numerical convergence. Recently, Zhu {\it et. al.} developed a direct SIM solution for the scattering of a circular cylinder with the singularity subtraction of the Green's function where no iteration is required \cite{Zhu2016}; it was further extended by Guan {\it et. al.} to multiple cylindrical layers \cite{Guan2020}. To our knowledge, however, the SIM has not been available for multiple objects.

In this work, we extend the SIM to analyze the scattering of multiple circular PEC cylinders.  (Note that a similar problem in elastic waves has been treated by using the surface integral equation and addition theorem in \cite{Lee2011}; it may also use the SIM developed in this work.) We first derive the coupled surface integral equations for $P$ non-overlapping circular PEC cylinders. Through the addition theorem and Fourier series, we can rewrite the coupled integral equations into a set of algebraic equations in the spectral domain, which is then solved easily by the stabilized biconjugate gradient (BiCGSTAB) method \cite{BiCGSTAB} with an effective preconditioner.  We show that this SIM needs a sampling density (SD) of only 2-3 points per wavelength (PPW) to reach the 99\% engineering accuracy for the electric current density on the surfaces of the PEC cylinders. The new contribution of this work is the extension of the spectral integral method to an arbitrary number of circular PEC cylinders by using the fast Fourier transform and singularity subtraction, thus allowing a low sampling density close to the Nyquist rate of two points per wavelength. This new contribution paves the way to the future application of SIM to multiple cylinders with inhomogeneous media through its combination with the finite element method (FEM) or spectral element method (SEM) for the interior regions, as shown for the multiple layers of circular cylinders with inhomogeneities in \cite{Guan2021}.

The organization of the paper is as follows: In Section
\ref{Theory}, we present the detail formulations of the
SIM. Numerical results are shown in Section \ref{ResultsDiscussions}. Finally, concluding remarks
are given in Section \ref{conclusions}.

\section{Theory}\label{Theory}
\thispagestyle{plain}

Consider $P$ circular PEC cylinders of radii $a^p$ and surface $S^p$ centered at ${\v O}^p=(\rho_p,\phi_p)$ ($p=1,2,\cdots,P$) in a global polar coordinate system with its origin at ${\v O}^0$. The infinite background domain outside the PEC surfaces
\begin{equation}
S=\bigcup\limits_{p=1}^{P} S^p
\label{eq_S}
\end{equation}
is filled with a homogeneous medium with permittivity $\epsilon=\epsilon_{r}\epsilon_0$ and permeability $\mu=\mu_{r}\mu_0$.  The objective of this work is to develop a fast method to solve for the scattered electromagnetic fields everywhere in space.

For a TM$_z$ incident electric field ${\v E}^{inc}=\hat zE_z^{inc}$, the scattered electric field from these PEC cylinders can be written as
\begin{equation}
E_z^{sct}=-\hat z j\omega\mu\int\limits_{S} g({\vg\rho}-{\vg\rho}')J_z(\vg\rho')dS(\vg\rho')
\end{equation}
where $\J=\hat zJ_z=\hat z\bigcup\limits_{p=1}^{P}J_z^p=\bigcup\limits_{p=1}^{P} \hat \rho^p\times \H|_{S^p}$ is the induced electric current density on the PEC cylinder surfaces, $\hat \rho^p$ is the unit outward radial vector in the $p$-th cylinder's local coordinates, $\H$ is the magnetic field, and
\begin{equation}
g({\vg\rho}-{\vg\rho}')={1\over 4j}H_0^{(2)}(k|{\vg\rho}-{\vg\rho}'|)
\end{equation}
is the two-dimensional Green's function of the background medium with the wavenumber $k=\omega\sqrt{\mu\epsilon}$.

As the tangential electric field $E_z=E_z^{inc}+E_z^{sct}$ on the surfaces of all PEC cylinders must vanish, we have the electric field integral equation
\begin{equation}
\begin{aligned}
E_z^{sct}(\vg\rho)&=-{\omega\mu\over 4}\int\limits_{S} H_0^{(2)}(kR)J_z(\vg\rho')dS(\vg\rho')\\
&=-E_z^{inc}(\vg\rho), {\rm\, for\, }\vg\rho\in S
\label{eq_IE1}
\end{aligned}
\end{equation}
where $R=|\v R|$, $\v R={\vg\rho}-{\vg\rho}'$.  According to (\ref{eq_S}), equation (\ref{eq_IE1}) consists of the following $P$ coupled electric field integral equations:
\begin{equation}
\begin{aligned}
E_z^{sct}(\vg\rho)&=-{\omega\mu\over 4}\sum\limits_{p=1}^P\int\limits_{S^p} H_0^{(2)}(kR)J_z^p(\vg\rho')dS(\vg\rho')\\
&=-E_z^{inc}(\vg\rho), {\rm\, for\, }\vg\rho\in S^q, \quad q=1,\cdots,P
\end{aligned}
\label{eq_IE1B}
\end{equation}
  To solve these $P$ coupled integral equations, we focus on the scattered electric field produced on the $q$-th cylinder by the current on the $p$-th cylinder
\begin{equation}
E_z^{sct,qp}(\vg\rho)=-{\omega\mu\over 4}\int\limits_{S^p} H_0^{(2)}(kR)J_z^p(\vg\rho')dS(\vg\rho'), \vg\rho\in S^q
\label{eq_IE2}
\end{equation}
There are two situations depending on the values of $p$ and $q$: the self interaction $p=q$ and mutual interactions $p\ne q$. Below we discuss their treatments within the SIM.

\subsection{Self Interaction in the $p$-th Local Coordinate System}

For $p=q$, we have the self interaction in (\ref{eq_IE2}) that describes the scattered field on the $p$-th cylinder caused by its own induced current.  Naturally, it is more convenient to use the $p$-th local polar coordinate system to express the above equation (\ref{eq_IE2}).  This local coordinate system is centered at ${\v O}^p$ of the $p$-th cylinder, so we define the local polar position vector for the source on $S^p$ as ${\vg\rho}'^{p}=(\rho'^{p}=a^p,\phi'^{p})$, and that for the receiver on $S^p$ as ${\vg\rho}^p=(a^p,\phi^p)$.  Thus the self-induced scattered field on the $p$-th cylinder is
\begin{eqnarray}
E_z^{sct,pp}(a^p,\phi^p)=-{\omega\mu a^p\over 4}\int\limits_{0}^{2\pi} H_0^{(2)}(kR^{pp})J_z^p(a^p,\phi'^p)d\phi'^p
\label{eq_IE2A}
\end{eqnarray}
where for both the source and receiver on the $p$-th cylinder
\begin{equation}
R^{pp}=|\vg\rho^p-\vg\rho'^p|=2a^p|\sin{\phi^p-\phi'^p\over 2}|
\end{equation}
and the integral kernel $H_0^{(2)}(kR^{pp})$ is weakly singular because $R^{pp}$ can be zero.
The treatment of this weakly singular problem using FFT has been solved by the previous SIM for a single PEC cylinder in \cite{Zhu2016}, so we will skip its detail and only summarize the final result below.

Define the Fourier series and its Fourier coefficient
\begin{eqnarray}
J_z^p(a^p,\phi'^{p})&=&\sum\limits_{m=-\infty}^{\infty}\tilde j_m^pe^{jm\phi'^{p}}\cr
\tilde j_m^p&=&{1\over 2\pi}\int\limits_{0}^{2\pi}
J_z^p(a^p,\phi'^{p})e^{-jm\phi'^{p}}d\phi'^p
\label{eq_FS1}
\end{eqnarray}
for the induced electric current, and
\begin{eqnarray}
\hskip-1.0cm E_z^{sct,qp}(a^q,\phi^q)&=&\sum\limits_{n=-\infty}^{\infty}\tilde e_n^{sct,qp}e^{jn\phi^q}\cr
\tilde e_n^{sct,qp}&=&{1\over 2\pi}\int\limits_{0}^{2\pi} E_z^{sct,qp}(a^q,\phi^q)e^{-jn\phi^q}d\phi^q
\label{eq_FS2}
\end{eqnarray}
for the scattered field, respectively. Then, from the well-known convolution theorem, the scattered field on $S^p$ due to $J_z^p$ can be written explicitly in the spectral domain (i.e., Fourier domain) from (\ref{eq_IE2A}) as
\begin{equation}
\tilde e_n^{sct,pp}=-{\pi\omega\mu a^p\over 2}\tilde h_n^{pp}\tilde j_n^p
\label{eq_IE2B}
\end{equation}
where $\tilde h_n^{pp}$ is the Fourier coefficient of $H_0^{(2)}(kR^{pp})$, which is given in detail in \cite{Zhu2016} with the singularity subtraction technique for treating the weak singularity of the Hankel function; such a singularity subtraction \cite{Hu1995} is essential for the fast convergence of the SIM. Note that the weak singularity exists only for the self interaction, and not for the mutual interactions discussed below because $R^{qp}\ne 0$ when $q\ne p$.
\subsection{Mutual Interactions and Coordinate Translation}
\thispagestyle{plain}

 Now we focus on the mutual interaction case where $p\ne q$. In this case we define $\rho_>=\max(\rho,\rho')$ and $\rho_<=\min(\rho,\rho')$, so from the addition theorem (\ref{A01}) in Appendix we can rewrite (\ref{eq_IE2}) as
\begin{equation}
\begin{aligned}
E_z^{sct,qp}(\vg\rho)&=-{\omega\mu\over 4}\sum\limits_{m=-\infty}^{\infty}\int\limits_{S^p} H_m^{(2)}(k\rho_>)J_m(k\rho_<)\\
&\cdot e^{jm(\phi-\phi')}J_z^p(\vg\rho')dS(\vg\rho'), {\rm\, for\, }\vg\rho\in S^q
\end{aligned}
\label{eq_IE3}
\end{equation}
in the global coordinate system.  Alternatively, it is more convenient to write
equation (\ref{eq_IE3}) in the $p$-th local coordinate system centered at ${\v O}^p$ as
\begin{equation}
\begin{aligned}
E_z^{sct,qp}(\vg\rho^p)&=-{\omega\mu\over 4}\sum\limits_{m=-\infty}^{\infty}\int\limits_{S^p} H_m^{(2)}(k\rho^p)J_m(k\rho'^{p})\\
&\cdot e^{jm(\phi^p-\phi'^{p})}J_z^p(\vg\rho'^{p})dS(\vg\rho'^{p}), {\rm\, for\, }\vg\rho^p\in S^q
\label{eq_IE4}
\end{aligned}
\end{equation}
because for this case we know $\rho^p>\rho'^{p}$ as long as the cylinders do not overlap.  Note that this is particularly convenient for the integral over $S^p$ as $\vg\rho'^{p}=(a^p,\phi'^p)$ has a constant radius $a^p$.

On the other hand, $E_z^{sct,qp}(\vg\rho^p)$ written in the $p$-th local coordinates in (\ref{eq_IE4}) is not very convenient for an observation point on the $q$-th cylinder surface as $|\vg\rho^p|$ is not a constant when $q\ne p$.  However, we can use the $q$-th local coordinate system ${\vg\rho}^q$ to express the observation point by noting that
\begin{equation}
{\vg\rho}^p={\vg\rho}_{pq}+{\vg\rho}^q, \quad {\vg\rho}^q=(a^q,\phi^q)
\label{eq_pq}
\end{equation}
where ${\vg\rho}_{pq}={\v O}^q-{\v O}^p=(\rho_{pq},\phi_{pq})$ is the constant translation (displacement) vector from ${\v O}^p$ to ${\v O}^q$ (see Fig. 1).
Furthermore, as $\rho^q<\rho_{pq}$ because the $P$ cylinders do not overlap, from the addition theorem in (\ref{A06}) in Appendix we can rewrite $H_m^{(2)}(k\rho^p)$ in (\ref{eq_IE4}) as
\begin{eqnarray}
\hskip-0.5cm H_m^{(2)}(k\rho^p)e^{jm\phi^p}&=&\sum\limits_{n=-\infty}^{\infty} J_{n} (k\rho^q)e^{jn\phi^q}\cr
&&\quad \cdot H_{m-n}^{(2)}(k\rho_{pq})e^{j(m-n)\phi_{pq}}
\label{eq_Add}
\end{eqnarray}

Therefore, in the $q$-th coordinate system equation (\ref{eq_IE4}) becomes
\begin{eqnarray}
&&\hskip-0.9cm E_z^{sct,qp}(\vg\rho^q)={\omega\mu a^p\over 4}\sum\limits_{m,n=-\infty}^{\infty}\int\limits_{0}^{2\pi} J_m(ka^p)J_{n}
(ka^q)e^{j(m-n)\phi_{pq}}\cr
&&\hskip1.5cm \cdot H_{m-n}^{(2)}(k\rho_{pq})e^{jn\phi^q}e^{-jm\phi'^{p}}J_z^p(\phi'^{p})d\phi'^p
\label{eq_IE5}
\end{eqnarray}
With the Fourier series, equation (\ref{eq_IE5}) can be written in the spectral domain as
\begin{eqnarray}
\tilde e_n^{sct,qp}&=&{\pi\omega\mu a^p\over 2}\sum\limits_{m=-\infty}^{\infty} J_m(ka^p)J_{n}
(ka^q)e^{j(m-n)\phi_{pq}}\cr
&&\qquad \hskip 2cm \cdot H_{m-n}^{(2)}(k\rho_{pq})\tilde j_m^p
\label{eq_IE6}
\end{eqnarray}
This completes the spectral domain formulation for the mutual interactions between the $p$-th and $q$-th cylinders ($p\ne q$).

\subsection{System of Equations and Its Solution}

 Using (\ref{eq_IE2B}) and (\ref{eq_IE6}) in the integral equations (\ref{eq_IE1}) in the spectral domain, we obtain the following algebraic equations for the unknowns $\{\tilde j_m^p\}$ in the spectral domain
\begin{equation}
\hskip-0.05cm\sum\limits_{p=1}^{P}\sum\limits_{m=-\infty}^{\infty} Z_{nm}^{qp}\tilde j_m^p={2\over\pi\omega\mu}\tilde e_n^{inc,q},\,
\begin{aligned} &n=0,\pm 1,\pm 2,\cdots\cr
&q=1,2,\cdots, P
\end{aligned}
\label{eq_LS1}
\end{equation}
where
\begin{eqnarray}
Z_{nm}^{qp}=\left\{\begin{array}{ll}
a^q\tilde h_n^{qq}\delta_{mn}, &\mbox{for $p=q$}\cr
a^pJ_m(ka^p)J_{n}(ka^q)e^{j(m-n)\phi_{pq}}\cr
\quad\cdot H_{m-n}^{(2)}(k\rho_{pq}), & \mbox{for $p\ne q$}
\end{array}\right.
\label{eq_LS2}
\end{eqnarray}
and $\tilde e_n^{inc,q}$ is the Fourier series coefficient of $E_z^{inc}$ on the $q$-th cylinder surface. After we choose a sampling density, we can calculate the total number of sampling points on every circle. However, this number isn’t always an integer, and therefore, we could find the closest odd to the number as the real total number of sampling points of that circle $M_0$. And $M=(M_0-1)/2$.

Equation (\ref{eq_LS1}) has an infinite number of coupled equations, so its exact solution is difficult.  However, numerically it is unnecessary to carry out this infinite sum; usually the summation over $m$ is truncated to $-M\le m\le M$ (i.e., $M$ multipoles), and only $N=2M+1$ equations are kept for index $n$.  Then the truncated equation (\ref{eq_LS1}) consists of $PN$ unknowns $\{\tilde j_m^q\}$ in the system.

The SIM solves the truncated $PN$ equations in (\ref{eq_LS1}) by using the stabilized biconjugate-gradient (BiCGSTAB) in the spectral domain, where $\{\tilde j_m^p\}$ are solved. Here we can use the induced current densities in the isolated single PEC cylinders (i.e., no mutual interactions among the cylinders)
\begin{equation}
(\tilde j_m^p)^{(0)}=-{2\over\pi\omega\mu a^p\tilde h_m^{pp}}\tilde e_m^{inc,p}\equiv q_{m}^{p}\tilde e_m^{inc,p}
\label{eq_LS3}
\end{equation}
 as the initial solution, where $q_{m}^{p}=-2/[\pi\omega\mu a^p\tilde h_m^{pp}]$.  Furthermore, to speed up the convergence of the iterative solution, the diagonal matrix equivalent to the solution of isolated single PEC cylinders is used as the preconditioner.  Thus, the truncated equation (\ref{eq_LS1}) is now solved by
 \begin{equation}
{\bf Q}{\bf Z}{\tilde{\bf j}}={\bf Q}{\tilde{\bf e}}^{inc}
\label{eq_LS4}
\end{equation}
where the preconditioner is $Q_{nm}^{pq}=q_{m}^{p}\delta_{mn}\delta_{pq}$, $\tilde{\bf j}$ denotes the vector of ${\tilde j_m}^p$, and $\tilde{\bf e}^{inc}$ denotes the vector of $\tilde e_m^{inc,p}$.
  For all the numerical examples in the next section, the number of iterations is only less than 8 for the residual error to decrease to less than $10^{-6}$. Then the FFT algorithm is used to convert the spectral domain current density to the spatial domain. If $M$ multipoles are used in (\ref{eq_LS2}), the computational complexity is $O(KP^2N^2)$, where $K$ (usually less than 8) is the number of iterations.

\subsection{Far-Field Solution}
 It is often required to evaluate the far-zone field distributions.  For this purpose, we denote ${\bf O}^0=(0,0)$ corresponding to $q=0$ as the origin of the global coordinate system; thus ${\vg\rho}={\vg\rho}^0$.  By virtue of equation (\ref{A07}) in Appendix, on a circle with a radius $a^0$ large enough ($a^0\ge \rho_{p0}$ for all $p$) to enclose all the $P$ circular cylinders, the addition theorem reads
 \begin{eqnarray}%
&H^{(2)}_m(k\rho^p)e^{jm\phi^p} =
\sum\limits_{n=-\infty}^{\infty} J_n(k\rho_{p0})e^{jn\phi_{p0}}\notag\\
&\hskip4cm \cdot H^{(2)}_{m-n}(k\rho^0)e^{j(m-n)\phi^0}
\label{eq_FF0}
\end{eqnarray}
 Thus, from (\ref{eq_IE4}) we have the scattered field given by the sum of all contributions from these cylinders:
\begin{align}
E_z^{sct,0}(\vg\rho^0)=&\sum\limits_{p=1}^{P}{\omega\mu a^p\over 4}\sum\limits_{m,n=-\infty}^{\infty}\int\limits_{0}^{2\pi} J_m(ka^p)H_{m-n}^{(2)}
(ka^0)\notag \\
&\cdot e^{j(m-n)\phi^0}J_n(k\rho_{p0})e^{jn\phi_{p0}}e^{-jm\phi'^{p}}J_z^p(\phi'^{p})d\phi'^p\cr\notag \\
& \hskip-0.5cm =\sum\limits_{p=1}^{P}{\pi\omega\mu a^p\over 2}\sum\limits_{m,n=-\infty}^{\infty} J_m(ka^p)H_{m-n}^{(2)}
(ka^0)\notag \\
&\cdot e^{j(m-n)\phi^0}J_n(k\rho_{p0})e^{jn\phi_{p0}}\tilde j_m^p
\label{eq_FF1}
\end{align}

When $a^0=|{\vg\rho}^0|\rightarrow\infty$, we can use the asymptotic form of Hankel functions in (\ref{eq_FF1}) to arrive at
\begin{align}
E_z^{sct,0}&\rightarrow\sqrt{2\over \pi ka^0}e^{-j(ka^0-{\pi\over 4})}\notag \sum\limits_{p=1}^{P}{\pi\omega\mu a^p\over 2}\notag \\ &\cdot\sum\limits_{m,n=-\infty}^{\infty} J_m(ka^p)J_n(k\rho_{p0})
e^{j(m-n)(\phi^0+{\pi\over 2})}e^{jn\phi_{p0}}\tilde j_m^p\cr\notag \\
&=\sqrt{\pi\omega\mu\eta\over 2 a^0}e^{-j(ka^0-{\pi\over 4})}
\sum\limits_{p=1}^{P}a^pA_p(\phi^0) B_p(\phi^0)
\label{eq_FF2}
\end{align}
where
\begin{eqnarray}
A_p(\phi^0)&=&\sum\limits_{m=-\infty}^{\infty}J_m(ka^p)e^{jm(\phi^0+{\pi\over 2})}\tilde j_m^p,\quad\notag\\
B_p(\phi^0)&=&\sum\limits_{n=-\infty}^{\infty}
 J_n(k\rho_{p0})e^{jn(\phi_{p0}-\phi^0-{\pi\over 2})}
\label{eq_FF3}
\end{eqnarray}
The radar cross section is thus given by
\begin{eqnarray}
\sigma_{2D}={\pi^2\omega\mu\eta\over |E_z^{inc}|^2}\Big|\sum\limits_{p=1}^{P}a^p A_p(\phi^0)B_p(\phi^0)\Big|^2
\label{eq_FF4}
\end{eqnarray}
where the incident electric field is that of a plane wave.  Note that the near field in (\ref{eq_FF1}) is written in a double summation over $m$ and $n$, while the far-field in (\ref{eq_FF2}) and the radar cross section are simplified to only two single summations $A_p$ and $B_p$ through (\ref{eq_FF3}).

\section{Results and Discussions}\label{ResultsDiscussions}
\thispagestyle{plain}

In this section, we show three numerical experiments to verify the accuracy and efficiency of the SIM. The first two examples compare the SIM and MoM results for the induced current density on the PEC surfaces; they verify that the developed SIM is more efficient and accurate for the EM scattering of multiple circular PEC cylinders. The third example shows the accuracy of the RCS solution from the SIM using the simplified far-field expression. The background medium of these three models is set as vacuum.
 The general configuration of our problem is shown in Fig. 1. Structures corresponding to every figure can be obtained by substituting the center positions and radii described in the corresponding caption into Fig. 1. A 100 MHz TM$_z$ plane wave of unit magnitude is incident along the $+x$ direction with its phase being zero at the origin, i.e., $E_z^{inc}=\exp(-j{2\pi\over 3}x)$. The computation is performed by MATLAB R2020a on a computer with the
system Microsoft Windows 10 Professional X64, 2019. The processor is Intel(R) Core(TM) i5-1035G4 CPU @1.10GHz, 1498 MHz and the memory is 16 GB.

\subsection{Example 1: Three Identical PEC Cylinders}

The first model consists of three circular PEC cylinders with the same radius $a^1=a^2=a^3=5$ m and centered at ${\v O}^1=(0,0)$, ${\v O}^2=(0,20)$ and ${\v O}^3=(35,21)$ m, respectively, in Cartesian coordinates.  SIM and MoM are used to calculate the induced current densities on these cylinders. The MoM used the 3-point Gaussian quadrature to calculate the integration of every segment. In order to obtain a solution with roughly the same accuracy, the SD is set as 30 PPW in MoM and 3 PPW in SIM (thus $M=15$ for all cylinders), as shown in Fig. 2(a). Because the current density in both methods is calculated at the sampling points, a different SD leads to different observation points. Therefore, in order to compare the two results, we adopt the sampling points of SIM as our common points for comparison and then use cubic spline interpolation to obtain the corresponding values of these points in the results calculated by MoM. The relative difference of the current density $errJ_{z,SM}$ is defined as
\begin{equation}
errJ_{z,SM}=\frac{\left\|J_{z,S}-J_{z,M}\right\|_2}{ \left\| J_{z,M}\right\|_2}
\end{equation}
where $J_{z,S}$ and $J_{z,M}$ denote the current density computed by SIM and MoM, respectively.

In this simulation, The 2-D MoM solver in FEKO requires 139.524 seconds of CPU time and 410.175MB memory, while the cost of SIM is 0.196 s and  24.03 MB, respectively. The relative difference between these two methods is 0.3\%, indicating that SIM is more effective and less memory consuming than MoM, and their results are in good agreement with each other.  Furthermore, the traditional low-order MoM requires a much higher SD to reach good accuracy.

\begin{figure}[H]
     \centering
     \begin{subfigure}[b]{0.44\textwidth}
         \centering
         \includegraphics[width=\textwidth]{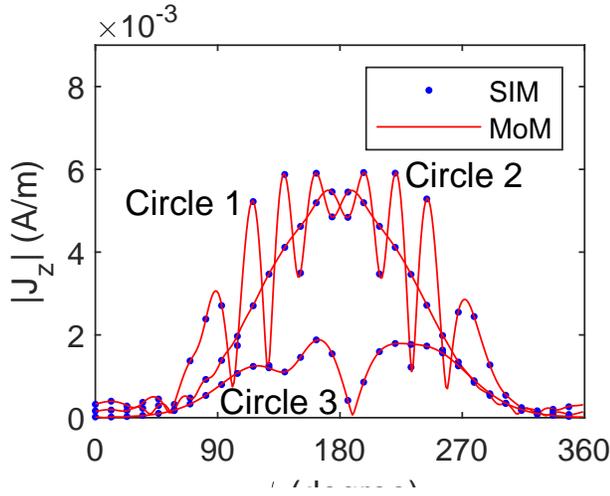}
         \caption{Comparison of current densities.}
         \label{Fig.2a}
     \end{subfigure}
     \hfill
     \begin{subfigure}[b]{0.44\textwidth}
         \centering
         \includegraphics[width=\textwidth]{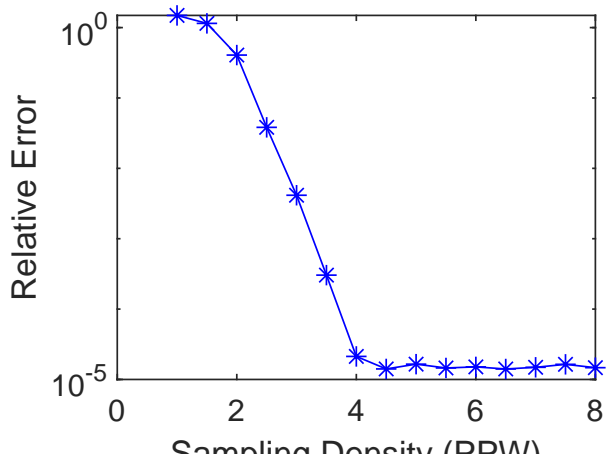}
         \caption{Error convergence. }
         \label{Fig.2b}
     \end{subfigure}
     \hfill
     \begin{subfigure}[b]{0.44\textwidth}
         \centering
         \includegraphics[width=\textwidth]{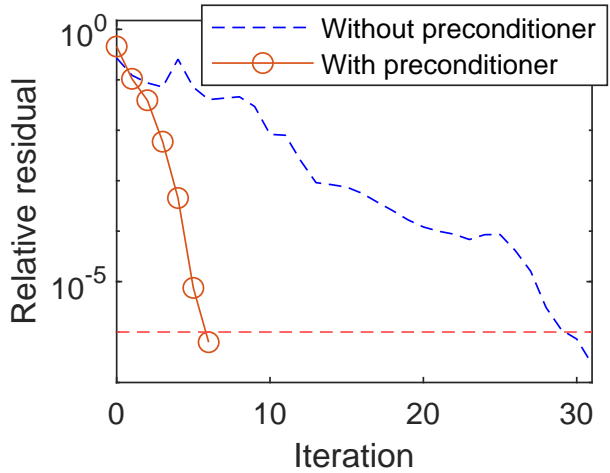}
         \caption{Preconditioner effect. }
         \label{Fig.2c}
     \end{subfigure}
        \caption{Example 1: Scattering of three circular PEC cylinders of the same radius of 5 m and centered at (0, 0), (0, 20) and (35, 21) m, respectively, in Cartesian coordinates. (a) $|J_z|$ on the PEC surface obtained by the SIM at 3 PPW and MoM at 30 PPW.  (b) Relative error of $J_z$ versus the sampling density in SIM. (c) Comparison of the relative residual error versus iteration between the methods with and without preconditioner.}
        \label{fig:three graphs}
\end{figure}

Based on the data above, we choose the SIM result at 20 PPW as the baseline reference to study the error convergence in SIM. Fig. 2(b) shows how the relative error of the SIM current density changes with the sampling density. It can be seen that it is an exponential convergence, and the SIM requires less than 3 PPW to reach 99\% accuracy.

To further demonstrate the effects of the preconditioning with the initial single-cylinder solutions, we show the error convergence curves for the cases with and without the preconditioning in Fig. 2(c).  It is shown that it takes only 6 iterations for the preconditioned solution to converge to a residual error of $10^{-6}$, compared to 32 iterations without preconditioning.

\subsection{Example 2: Five Different PEC Cylinders}
\thispagestyle{plain}
The second model consists of five PEC cylinders with different radii of 30, 18, 24, 12 and 36 m, and centered at positions ${\v O}^1=(0,-100)$, ${\v O}^2=(0,200)$, ${\v O}^3=(350,210)$, ${\v O}^4=(500,170)$ and ${\v O}^5=(-250,120)$ m, respectively, in Cartesian coordinates.

\begin{figure}[h]
     \centering
     \begin{subfigure}[b]{0.45\textwidth}
         \centering
         \includegraphics[width=\textwidth]{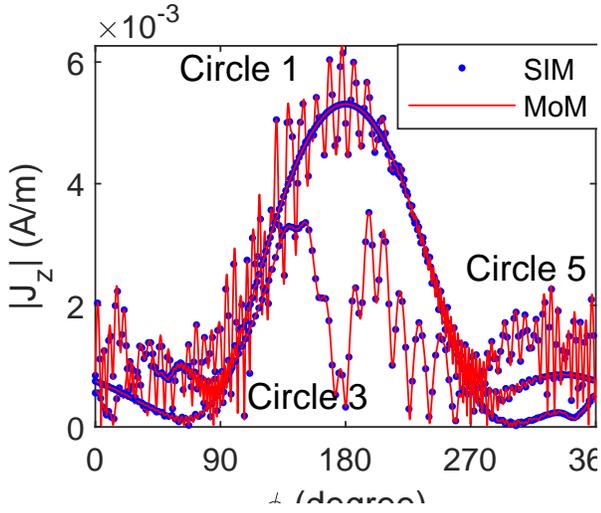}
         \caption{Comparison of current densities.}
         \label{Fig.3a}
     \end{subfigure}
     \hfill
     \begin{subfigure}[b]{0.45\textwidth}
         \centering
         \includegraphics[width=\textwidth]{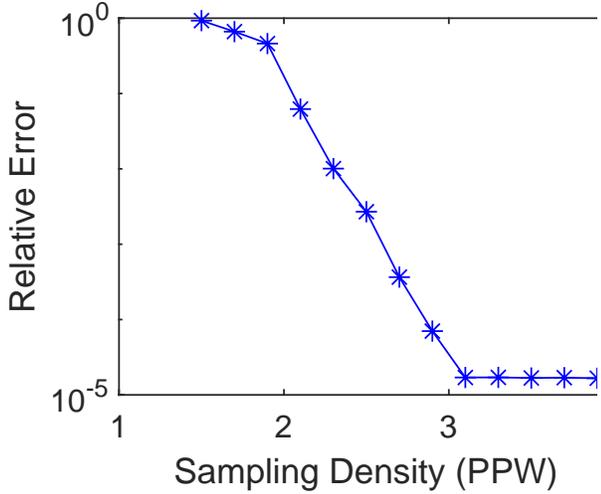}
         \caption{Error convergence. }
         \label{Fig.3b}
     \end{subfigure}

        \caption{Example 2: Scattering of five different PEC circular cylinders or radii 30, 18, 24, 12 and 36 m, centered at (0,-100), (0,200), (350,210), (500,170) and (-250,120) m, respectively, in Cartesian coordinates. (a) $|J_z|$ on three of the five circular PEC cylinder surfaces obtained by the SIM at 3 PPW and MoM at 30 PPW.  (b) Relative error of $J_z$ versus the sampling density in SIM.}
        \label{Fig}
\end{figure}

Similar to the first model, the SIM and MoM are both used to calculate the induced current densities. The sampling density is set as 3 PPW for SIM and 30 PPW for MoM, respectively to maintain consistency in accuracy (thus $M=94, 56, 75, 37, 113$ for the five cylinders, respectively). For clarity, only the current density on three of the five cylinders is shown in Fig. 3(a).
 The 2-D MoM solver in FEKO requires a much larger memory than that on our desktop computer. Therefore, we just used our own MoM code, which requires 40187.89 s of CPU time and 8.38 GB memory,
 while the cost of SIM is 0.847 s and 0.20 GB, respectively. The relative difference between the two methods is only 0.3\%. This result indicates that SIM is much more efficient than MoM and requires a much lower memory with approximately the same accuracy.

Based on the data above, we still choose the SIM result at 20 PPW as the baseline reference for studying the convergence of SIM. The error convergence curve of SIM is also exponential, as shown in Fig. 3(b). It can be seen that good accuracy of better than 99\% can be achieved with a SD between 2 and 3 PPW. This indicates that the convergence rate of SIM is very fast.
Through this study, we can see that SIM saves 90\% sampling points compared with MoM when computing current densities. Moreover, benefiting from its much fewer sampling points, the semi-analytical method and characteristics of SIM itself, the proposed SIM is much more efficient than MoM.

\subsection{Example 3: RCS of Five Identical PEC Cylinders}

The above near field obtained by the SIM can be used to calculate the field everywhere, including the far field and radar cross section (RCS). Here we use the simplified formulas in (\ref{eq_FF4}) to obtain the RCS. The model used in this simulation consists of five PEC cylinders with the same radius $a^1=a^2=a^3=a^4=a^5=6$ m, centered at positions ${\v O}^1=(0,-10)$, ${\v O}^2=(0,20)$, ${\v O}^3=(35,21)$, ${\v O}^4=(50,17)$ and ${\v O}^5=(-25,12)$ m, respectively, in Cartesian coordinates. The commercial integral equation solver FEKO with 2-D MoM (under the $z$-direction periodic boundary condition setting in 3D) is used as a reference.  Note that FEKO apparently is unable to solve such a special 3-D problem with the $z$-direction periodic boundary condition with the multilevel fast multipole method.  The SD is set as 3 PPW for SIM (i.e., $M=19$ for all cylinders), and the mesh is set as ``fine'' in FEKO. It is shown in Fig. 4(a) that the current densities computed by SIM and FEKO are in good agreement: the relative difference between them is only 0.75\%. The bistatic RCS curves are obtained over the 360 degrees equally divided by 720 observation angles, as shown in Fig. 4(b). It can be observed that good agreement has been achieved, and the relative difference between the two solvers is only 0.924\% for RCS. In this case, the SIM takes 0.555 s, while FEKO takes 389.789 s, so the SIM is 701 times faster than the 2-D MoM in FEKO.

\begin{figure}[h]
    \centering
    \begin{subfigure}[b]{0.45\textwidth}
         \centering
         \includegraphics[width=\textwidth]{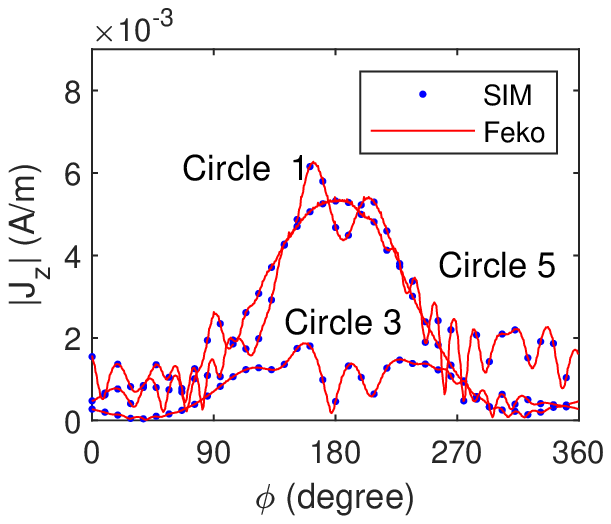}
         \caption{Comparison of current densities.}
         \label{Fig.4a}
     \end{subfigure}
     \hfill
     \begin{subfigure}[b]{0.45\textwidth}
         \centering
         \includegraphics[width=\textwidth]{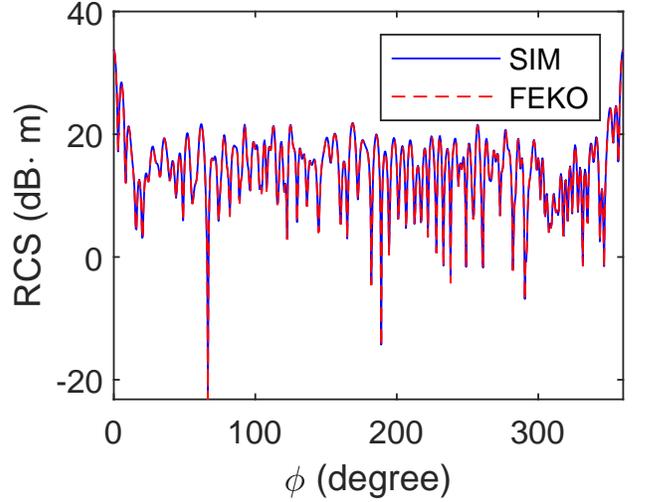}
         \caption{Comparison of RCS.}
         \label{Fig.4b}
     \end{subfigure}

        \caption{Example 3: Comparison of current densities for the scattering of five circular PEC cylinders of the same radius of 6 m centered at (0,-10), (0,20), (35,21), (50,17) and (-25,12) m, respectively, in Cartesian coordinates.  (a) $|J_z|$ on the first, third and fifth cylinders.  (b) The bistatic RCS obtained by SIM and FEKO.}
        \label{Fig.4}

\end{figure}
\thispagestyle{plain}

\section{Conclusions}\label{conclusions}

We have developed a spectral integral method for the scattering of an arbitrary number of non-overlapping circular PEC cylinders through the fast Fourier transform and addition theorem.  It is an extension of the previous SIM for a single object. It is shown that this SIM can achieve very accurate results with a sampling density of only slightly above 2 points per wavelength.  Numerically, to achieve an accuracy of 99\%, it can be 100 times more efficient and 6 times less memory-consuming than the MoM in the simulated cases.  Thus, it provides an efficient and highly accurate technique to solve the scattering problem of multiple circular PEC cylinders.  Furthermore, this SIM can be further extended to the cases of multiple dielectric cylinders.  More importantly, using the SIM as an exact radiation boundary condition, we can combine it with the finite element method and spectral element method to obtain highly efficient solutions for multiple objects with arbitrary inhomogeneity and anisotropy. These will be part of our future work.

\appendix[Addition Theorem]
\thispagestyle{plain}

The basic addition theorem is well known for the zeroth order cylindrical harmonics, for example:
\begin{equation}
H_0^{(2)}(k|\vg{\rho}-\vg{\rho}'|) =\left\{\begin{array}{ll}
\sum\limits_{n=-\infty}^{\infty} J_n
(k\rho)H_n^{(2)}(k\rho ')\\
 \qquad \cdot e^{jn(\phi-\phi')}  \quad\mbox{for $\rho \leq \rho'$}\\
\sum\limits_{n=-\infty}^{\infty} J_n
(k\rho')H_n^{(2)}(k\rho)\\
\qquad \cdot e^{jn(\phi-\phi')} \quad\mbox{for $\rho \geq \rho '$}
\end{array}\right.
\label{A01}
\end{equation}
where $\vg{\rho}=(\rho,\phi)$ and $\vg{\rho}'=(\rho',\phi')$ are the global polar coordinates of the observer and source, respectively.  The displacement vector is
\begin{equation}
{\v R}=\vg{\rho}-\vg{\rho}'=(R,\phi_R)
\label{A02}
\end{equation}
in the global polar coordinates.

Similarly, for higher order cylindrical harmonics, we have \cite{Watson}
\begin{equation}
{\cal C}_m(kR)e^{jm(\phi-\phi_R)} =\left\{\begin{array}{ll}
\sum\limits_{n=-\infty}^{\infty} J_n
(k\rho){\cal C}_{m-n}(k\rho ')\\
\quad\cdot e^{jn(\phi-\phi'-\pi)} \quad \mbox{for $\rho \leq \rho'$}\\
\sum\limits_{n=-\infty}^{\infty} J_n
(k\rho'){\cal C}_{m-n}(k\rho)\\
\quad\cdot e^{jn(\phi-\phi'-\pi)} \quad \mbox{for $\rho \geq \rho '$}
\end{array}\right.
\label{A03}
\end{equation}
where $m$ can be any real number, and ${\cal C}_m$ represents any Bessel and Hankel functions $J_m$, $Y_m$, $H_m^{(1)}$ and $H_m^{(2)}$.

Now suppose we have two local coordinate systems with their origins located at ${\v O}^p$ and ${\v O}^q$, respectively.  The displacement from ${\v O}^p$ to ${\v O}^q$ is defined as
\begin{equation}
{\vg\rho}_{pq}={\v O}^q-{\v O}^p=(\rho_{pq},\phi_{pq})
\label{A04}
\end{equation}
in the global polar coordinates.  For a position vector $\vg{\rho}$, if the local polar coordinates in systems ${\v O}^p$ and ${\v O}^q$ are given by
\begin{equation}
{\vg \rho}^p=(\rho^p,\phi^p), \quad {\vg\rho}^q=(\rho^q,\phi^q)
\label{A05}
\end{equation}
respectively, then we have the following addition theorem:
\begin{equation}
\hskip-0.3cm {\cal C}_m(k\rho^p)e^{jm(\phi^p-\phi_{pq})} =\left\{\begin{array}{ll}
\sum\limits_{n=-\infty}^{\infty} J_n
(k\rho^q){\cal C}_{m-n}(k\rho_{pq})\\
\cdot e^{jn(\phi^q-\phi_{pq})} \,\mbox{ for $\rho^q \leq \rho_{pq}$}\\
\sum\limits_{n=-\infty}^{\infty} {\cal C}_{n}(k\rho^q)J_{m-n}
(k\rho_{pq})\\
\cdot e^{jn(\phi^q-\phi_{pq})} \,\mbox{ for $\rho^q \geq \rho_{pq}$}
\end{array}\right.
\label{A06}
\end{equation}
Alternatively, we can write this as
\begin{eqnarray}
\hskip0cm {\cal C}_m(k\rho^p)e^{jm\phi^p} =\left\{\begin{array}{ll}
\sum\limits_{n=-\infty}^{\infty} J_n(k\rho^q)e^{jn\phi^q}{\cal C}_{m-n}(k\rho_{pq})\\
\quad\cdot e^{j(m-n)\phi_{pq}}
\mbox{ for $\rho^q \leq \rho_{pq}$}\\
\sum\limits_{n=-\infty}^{\infty} J_n(k\rho_{pq})e^{jn\phi_{pq}}{\cal C}_{m-n}(k\rho^q)\\
\quad\cdot e^{j(m-n)\phi^q}
\mbox{ for $\rho^q \geq \rho_{pq}$}
\end{array}\right.
\label{A07}
\end{eqnarray}
This effectively transforms the cylindrical harmonics in ${\v O}^p$ coordinate system to those in ${\v O}^q$ coordinate system.
\newpage

\bibliographystyle{IEEEtran}

\end{document}